# Polarisable force fields: What do they add in biomolecular simulations?

*V. S. Sandeep Inakollu,[1,2,3] Daan P. Geerke,[4,*] Christopher N. Rowley,[5,*] Haibo Yu[1,2,3,*]*

1. School of Chemistry and Molecular Bioscience, University of Wollongong, Wollongong NSW 2522, Australia

2. Molecular Horizons, University of Wollongong, Wollongong NSW 2522 Australia

3. Illawarra Health and Medical Research Institute, Wollongong, New South Wales, 2522, Australia

4. AIMMS Division of Molecular and Computational Toxicology, Department of Chemistry and Pharmaceutical Sciences, Vrije Universiteit Amsterdam, De Boelelaan 1108, 1081 HZ Amsterdam, the Netherlands

5. Department of Chemistry, Memorial University of Newfoundland, St. John's, Newfoundland and Labrador, Canada

*Corresponding authors: d.p.geerke@vu.nl, crowley@mun.ca, hyu@uow.edu.au

## Abstract

The quality of biomolecular simulations critically depends on the accuracy of the force field used to calculate the potential energy of the molecular configurations. Currently, most simulations employ non-polarisable force fields, which describe electrostatic interactions as the sum of Coulombic interactions between fixed atomic charges. Polarisation of these charge distributions is incorporated only in a mean-field manner. In the past decade, extensive efforts have been devoted to developing simple, efficient, and yet generally applicable polarisable force fields for biomolecular simulations. In this review, we summarise the latest developments in accounting for key biomolecular interactions with polarisable force fields and applications to address challenging biological questions. In the end, we provide an outlook for future development in polarisable force fields.

## Introduction

Atomistic modelling plays an increasingly important role in understanding the structure-function-dynamics relationship in biomolecular systems. This understanding now facilitates various types of molecular engineering that would have been impossible without the insights provided by modelling [1]. The accuracy and predictive power of molecular dynamics (MD) simulations based on all-atom force fields are steadily improving due to the parallel improvements in high-performance computing hardware, more accurate methods for calculating the potential energy of a conformation, and more efficient methods for conformational sampling. Nowadays, μs-length simulations of systems containing hundreds of thousands of atoms are performed routinely. With specialised supercomputers, it has been possible to perform millisecond-length simulations,[2] and the simulations of entire cellular structures have been attempted.[3]

The general form of the widely-used conventional force fields dates back to the pioneering work by Lifson's group.[4,5] It consists of the bonded interactions (bonds, valence angles, dihedral angles) and the nonbonded interactions (both electrostatic and van der Waals). The van der Waals term is often described by a Lennard-Jones form, and the electrostatic interactions are described using Coulomb's law, with fixed partial charges preassigned to each atom according to the adopted force field. This type of force field is called an additive or non-polarisable force field. Force field developers have a variety of strategies to parameterise the partial charges.[6,7] One common feature among them is that the polarisation effect is treated in a mean-field manner, in which the partial charges and dipole moments are enhanced compared to their gas-phase values, mimicking the effect of induced polarisation in an average way. Although this model is simple, they have benefited from almost 40 years of parameterization refinements, and they have provided a wealth of information into complex molecular systems.[1] The inherent limitation of these models is that they are incapable of describing the change of polarisation of molecules when they adopt different conformations or encounter different interacting partners over the course of a simulation. For example, the polarisation of a solute is expected to increase when it moves from a non-polar region of the system into a polar region, but this effect is neglected by conventional non-polarisable models.

Developing computational models that account for induced polarisation has been a longstanding objective in computational biophysics.[8] However, the broad adoption of polarisable force fields in biomolecular simulations was hampered by the limited availability of model parameters and the increased computational cost. In recent years, there have been sustained efforts by several groups towards devising and parameterizing polarisable force fields for biomacromolecules. At the same time, the development of high-performance computing has allowed sufficient conformational sampling of systems of biological interest using these polarisable models[9,10]. For instance, using NAMD, the computational effort required for an MD simulation using a polarisable model is roughly double that of the non-polarisable counterpart, making these simulations tractable if sufficient computing resources are available[10].

There are at least three different methods to account for explicit polarisation in classical force fields:[11] the Point-Polarisable Dipole (PPD)[12,13], Fluctuation Charge (FQ)[14,15] and Drude Oscillator (DO) [16] (or called Shell Model[17], and Charge-on-Spring model[18]). Combined models can be found in the literature too. Huang et al. recently demonstrated that it is possible to map the electrostatic model optimised in the Drude force field onto the multipole and induced dipole model and illustrated the equivalency between DO and PPD.[19] This review article will focus on the latest developments in and applications of polarisable force fields for biomolecular simulations. We will not add extensive general references to various polarisable models, and readers are referred to the latest review articles.[20–22] First, we briefly review the recent development in dealing with challenging molecular interactions and highlight some of the latest applications of polarisable force fields. Finally, we present a summary and outlook.

# Fundamental key interactions

Additive force fields are the most commonly used force fields in biomolecular simulations. However, their accuracy can be limited by their use of fixed atomic charges. This is particularly significant for modelling processes where electrostatic interactions are changing and fluctuating or where induced polarisation is an essential part of the interactions. Compared to additive models, explicitly accounting for polarisation can increase the transferability of force field parameter sets in terms of their accuracy to describe intermolecular interactions in environments of different polarities. [23] Consequently, it is challenging to describe some key biomolecular interactions using additive models, such as cation–$\pi$ and metal/molecular–ion interactions. As described below, recent efforts have focused on developing polarisable force fields to describe such interactions accurately **(Figure 1)**. Moreover, the deficiencies in the models currently used to describe London dispersion interactions are noted.

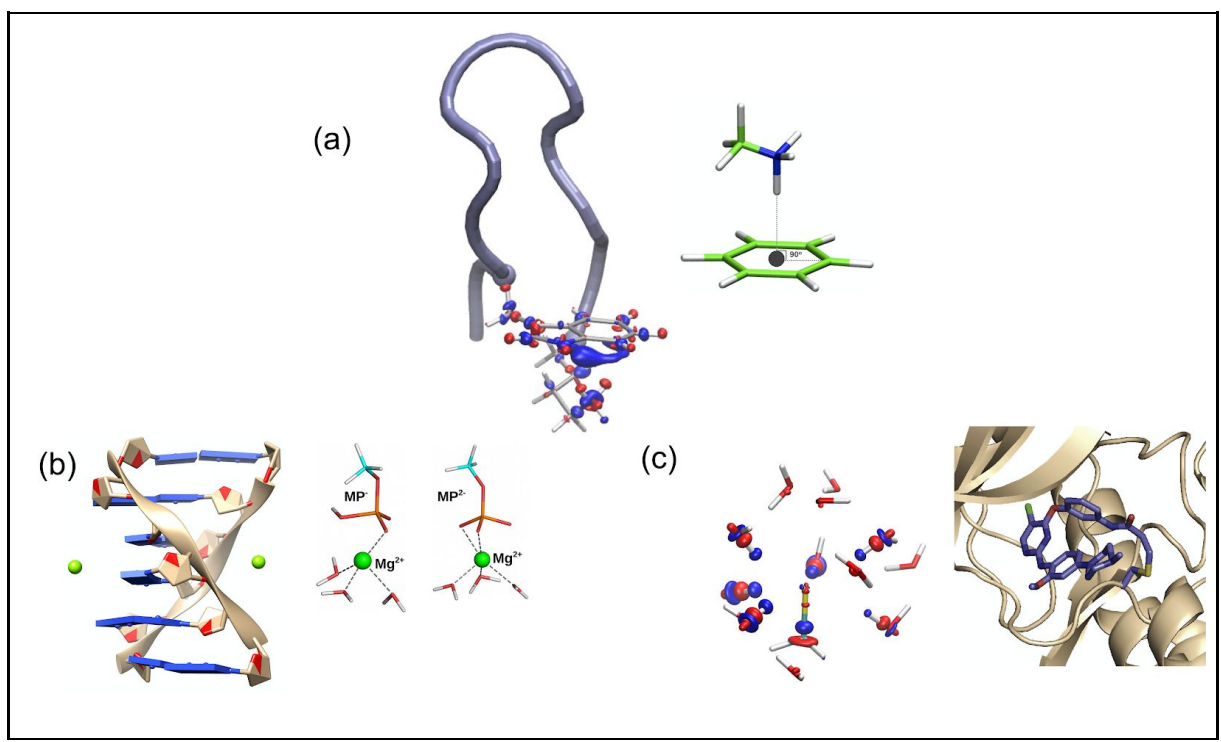

**Figure 1:** (a) Cation-π interactions in the biomolecular system illustrated by a cation-π interaction between Lys1 and Trp10 in the HP peptide (PDB ID: 2EVQ). The difference in electron density distributions between the interacting and non-interacting states (left) shows that the Trp π-electron density is polarised towards the cationic $NH_3^+$ group of the lysine (blue) away from the atomic nuclei (red). In the CHARMM-Drude model (right) this type of cation-π interaction is approximated by Drude oscillators tethered to the non-hydrogen atoms and an additional charge at the centre of the π ring (black point). (b) Metal and molecular ion interactions illustrated by Z-DNA crystal with 2 $Mg^{2+}$ (PDB ID: 1LJX). The CHARMM-Drude model (right) accurately describes the interactions between $Mg^{2+}$ and phosphate groups of the nucleic acid (taken from Ref. [24]). (c) Other biologically relevant elements and functional groups illustrated by covalent-modifier ibrutinib bound to TgCDPK1 (PDB ID: 4IFG, taken from Ref.[25]). The electron densities of water molecules coordinated to a model thiolate are polarised by the anionic charge (left).

**Cation-π and π-π interactions**

Cation-π interactions commonly occur between the positively charged cations and negatively charged π electron-rich cloud in the aromatic ring in the charged and aromatic amino acid or nucleic acids.[26] These interactions are highly anisotropic in nature. The polarisation and the charge redistributions are essential to model these interactions correctly. Rupakheti et al.[27] studied the commonly occurring cation-π interactions in the proteins between the aromatic and charged amino acids, by comparing the potentials of mean force (PMF) for a series of prototypical cation-π models with both CHARMM36 (C36) and the Drude-2013 polarisable force field.[28] Based on the reversible association PMFs, they showed that explicitly accounting for polarisation globally enhanced the description of the cation-π interactions. They also noted the challenges in accurately describing the interactions responsible for amino acid cation-π interactions. Lin and MacKerell[29] systematically optimised the CHARMM Drude-2013 polarisable force field parameters[28] for cation-π and anion-aromatic ring interactions, targeting the QM interaction energies and geometries. The atom pair-specific Lennard-Jones parameters along with virtual particles as selected ring centroids were introduced. The refined CHARMM Drude-2013 protein force field has been shown to provide a significant improvement in reproducing the ion-π pair distances observed in experimental protein structures **(Figure 2)**. Zhang et al.[30] developed the AMOEBA polarisable force field for aromatic molecules and nucleobases, in which their parameters were parameterised against the properties in the gas phase with QM calculations and experimental values in the condensed phase. They further extend the development to a full set of AMOEBA force fields for nucleic acids.[31]

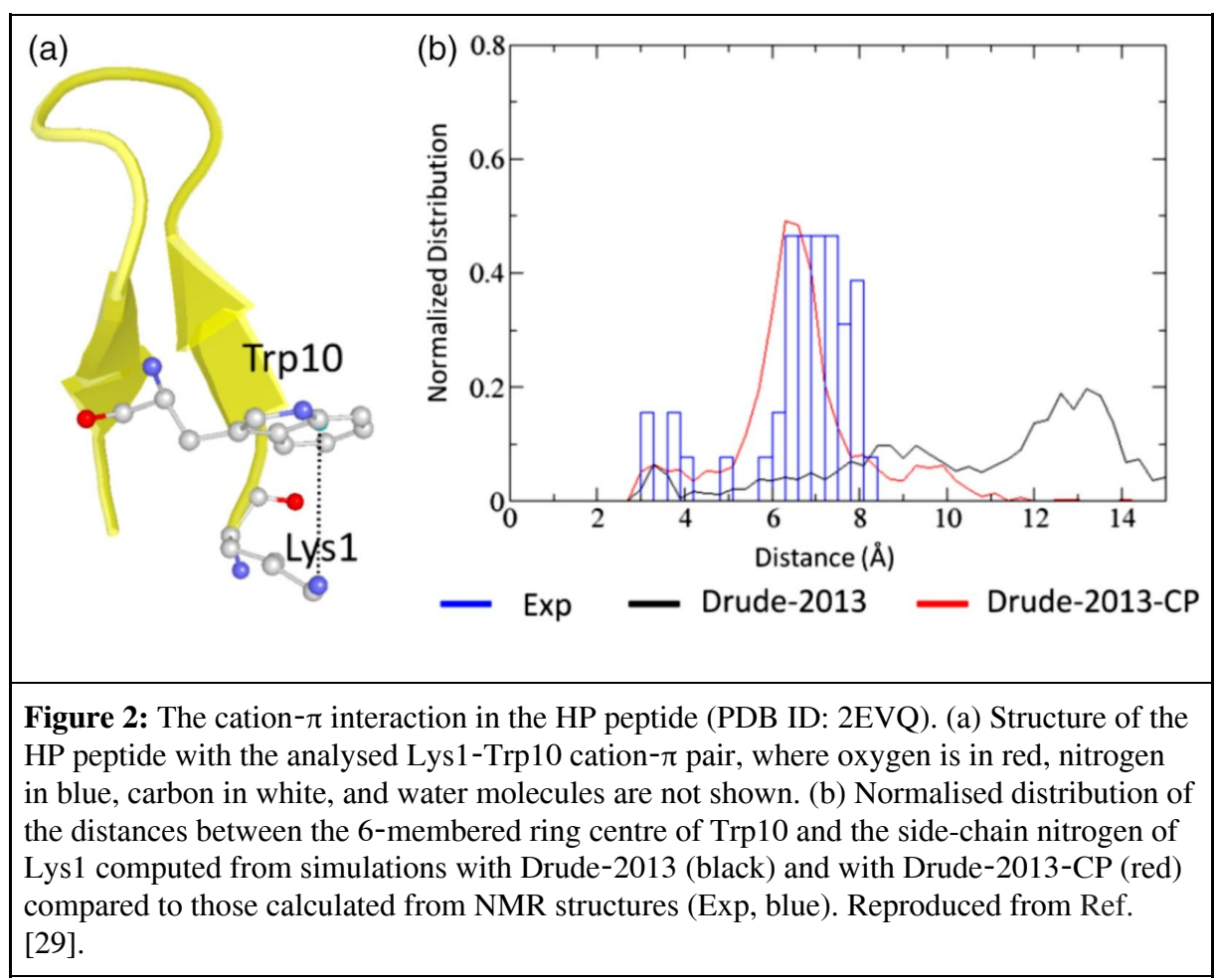

**Figure 2:** The cation-π interaction in the HP peptide (PDB ID: 2EVQ). (a) Structure of the HP peptide with the analysed Lys1-Trp10 cation-π pair, where oxygen is in red, nitrogen in blue, carbon in white, and water molecules are not shown. (b) Normalised distribution of the distances between the 6-membered ring centre of Trp10 and the side-chain nitrogen of Lys1 computed from simulations with Drude-2013 (black) and with Drude-2013-CP (red) compared to those calculated from NMR structures (Exp, blue). Reproduced from Ref. [29].

**Metal and molecular ion interactions**

Metal ions are fundamental to the structure and function of many biological systems, where they may interact with solvent, proteins, membranes and nucleic acids. The presence of the metal ion strongly alters the local electrostatic environment. Several studies have pointed out the intrinsic limitations of additive force fields in studying metal ion interactions.[32,33] Parameters have been developed for the set of biologically relevant ions for both the Drude and AMOEBA force fields.[34,35] The AMOEBA force field was used to study the selectivity for $Ca^{2+}$ and $Mg^{2+}$ ions for various protein binding pockets. It was shown that unless polarisation was included, the smaller ion $Mg^{2+}$ is always favoured over the larger ion $Ca^{2+}$.[36] Another notable recent development includes polarisable models for biologically relevant molecular ions.[37] For instance, phosphate groups are essential components of nucleic acids. Their interactions with the surrounding solvents, metal ions, and proteins facilitate the binding and folding motions in the nucleic acids. Lemkul and MacKerell [38] and Villa et al. [24] studied the interactions of phosphate analogues, including dimethyl phosphate (DMP) and methyl phosphate (MP), with the $Mg^{2+}$ ion with the Drude polarisable force field. The $Mg^{2+}$-phosphate-binding free energies calculated using the Drude model have better agreement with the QM and experimental data. Furthermore, the refined complete set of Drude polarisable force field for DNA and RNA has been reported and validated.[39–41] Similar work has been carried out for the AMOEBA force field.[42]

While these models provide potential energy surfaces that are in reasonable agreement with QM results, energy decomposition analysis (EDA) has revealed that the relative magnitude of the components of the interaction energy of the polarisable MM and QM models can be very different. In this analysis, the charge-penetration (CP), charge-transfer (CT), dispersion, permanent electrostatic, and polarisation interactions in water–water, water–ion, and ion–protein model compounds were calculated using EDA of the DFT interaction energy with the absolutely localised molecular orbitals (ALMO) scheme and compared to the components of the AMOEBA interaction energy.[43,44] AMOEBA does not include CP and CT terms, but in water–water interactions, the 14-7 potential used to represent van der Waals interactions in the AMOEBA model partially compensated for these effects. This cancelation of error was less effective for water–halide, water–divalent cation, and $Ca^{2+}$-protein models, where the magnitudes of permanent electrostatic and polarisation interactions in the AMOEBA model deviated significantly from the EDA results. These studies serve to guide the future parametrisation of explicit functional forms for short-range contributions from CP and/or CT.[45–47]

**Other biologically important elements and groups**

Cysteine is a unique sulphur amino acid involved in various biological processes, including protein-ligand binding, catalytic reactions, and post-translational modifications. Due to the presence of the thiol group, which has a moderate pKa, cysteine can exist in its anionic form under physiological conditions. Non-polarisable force fields have limited success in describing the structure and hydration energies of these highly polarisable ions. Lin et al.'s development of a CHARMM-Drude model for polyatomic ions provided the first polarisable model for thiolates.[37] Williams and Rowley[48] showed that the Drude polarisable model predicted the structural and energetic properties of methylthiolate in good agreement with QM/MM MD simulations, while the conventional MM model overestimated its solvation free energy. Recently, Drude polarisable force field parameters have been developed for halogen-containing compounds, which will allow this model to be used to model the binding of halogenated drugs to protein targets.[49]

**van der Waals interactions**

Although these polarisable models account for the induction of an atomic dipole from the electric field created by the environment around the atom, the instantaneous-dipole—induced dipoles that give rise to the London dispersion interactions are not captured. The pairwise Lennard-Jones potential or a similar 14-7 potential has been adopted to account for Pauli repulsive and dispersion forces in the polarisable force fields. As the electrostatic components of these force fields have changed, the van der Waals parameters of conventional force fields are no longer appropriate, so new parameters have to be determined for use with the polarisable force fields. Typically, non-bonded parameters of polarisable models are still assigned empirically based on bulk physical properties of liquids. While polarisable force fields typically have static charges and dipole-moments that are closer to their gas-phase QM estimates than additive force fields, molecular dispersion $C_6$ parameters are typically too high. [50,51]. Recently, new methods have been developed to define dispersion parameters from quantum chemical calculations, which has the potential to simplify force field development and make the models more transferable.[52,53]

# Protein simulations

Protein structure and dynamics are other areas where induced polarisation is expected to have a significant effect. For example, when proteins fold to form $\alpha$-helices, the NH and C=O moieties of the amide backbone form strong hydrogen bonds. The polarisation of these bonds results in a cooperative effect, where the strength of the hydrogen bonds increases as the number of turns in the helix increases.[54] Likewise, the cooperativity of hydrogen bonds between polar side chains can stabilize the folded state of a protein. The accurate description of the relative stability and transition rates between unfolded/misfolded and folded states will likely require explicit treatment of induced polarisation.[55]

These issues are particularly relevant in the simulation of intrinsically disordered proteins (IDP). IDPs are involved in several pathological disorders, including cancer and neurodegenerative disorders[56]. IDPs are characterised by the lack of well-defined tertiary structure. Instead, they exist in an interconverting ensemble of conformations. The amino acid sequence in IDPs is enriched with polar and charged amino acids, and have relatively low numbers of hydrophobic amino acids, which are essential for protein core formation.[57] Both Amber and CHARMM additive force fields have recently been refined to provide a better description of IDPs, although their performance is inconsistent.[58,59] Treatment of explicit polarisation may be needed to model the diverse range of structure IDPs exist in.[57,60] Wang et al.[61] conducted a study to compare the performance of non-polarisable and polarisable force fields for protein structural refinement, protein folding, and simulating IDPs. They showed that the inclusion of explicit polarisation improves accuracy in protein structure refinement and the description of IDP conformational ensembles. This study also noted the difficulties for the polarisable force field to sample the native structures in the selected proteins. To address this limitation, future work is required to further refine the parameters. This may well comprise improving the description of dispersion, which was recently shown to be important for the simulation of IDPs.[58]

Water dynamics on the surface of proteins play a significant role in protein folding and unfolding. Ngo et. al.[62] studied the hydration free energies of amino acid side chains, protein-water and protein-protein interactions, and the hydrogen-bond lifetime with the CHARMM additive C36 and Drude polarisable force fields. The side chain hydration energies predicted by the CHARMM Drude force field are generally in better agreement with the experimental data than that of the C36 force field, except for the acidic amino acid side chains. The development of revised CHARMM-Drude parameters for molecular ions may help resolve this issue.[37] In the simulations with the CHARMM Drude force field, stronger interactions and longer-lived hydrogen bonds between the first hydration shell and the protein were observed. Furthermore, the first solvation shell prevents other waters from accessing the protein surface.

Hazel et al.[63] studied the folding free energy landscapes of C-terminal $\beta$-hairpin of the B1 domain of streptococcal protein G (GB1) using replica exchange umbrella sampling simulations with two non-polarisable force fields (C36 and C22*) and the CHARMM-Drude-2013 polarisable force field. Surprisingly, the C22* and CHARMM-Drude model agreed better with the experimental studies of GB1 folding, while C36 over stabilises the $\beta$-hairpin. Current literature suggests that more validation studies and continuous refinement of the polarisable force fields are needed for it to be widely applicable in simulating protein dynamics.

## Protein-ligand interactions

Electrostatic interactions can play a major role in protein–ligand and enzyme–substrate interactions. Often the protein binding sites and the enzyme active sites encompass a heterogeneous environment that can also include water molecules and metal ions. This presents challenges for additive force fields, particularly for highly-charged species. Qi et al. used the AMOEBA polarisable force field in designing inhibitors for fructose-bisphosphate aldolase A (ALDOA).[64] ALDOA converts fructose-1,6 bisphosphate (FDP) into

glyceraldehyde-3-phosphate (GAP) and dihydroxyacetone phosphate. Substrate-mimicking inhibitors for ALDOA are typically highly charged. The AMOEBA simulations were applied to model the binding of a series of naphthalene-2,6-diyl bisphosphate analogues and rank their relative binding free energies, which match experimental data well. Panel et al. [65] studied binding specificity between the PDZ domain and C-terminal peptides of its target proteins, which form the building blocks of eukaryotic signalling pathways. It was found that the additive force field AMBER ff99SB over-stabilises salt-bridge interactions and the Drude force field significantly reduced errors for those involving ionic mutations. This suggests that electronic polarisation can be crucial to describe ionic interactions in buried regions.

Welborn and Head-Gordon[66] used the AMOEBA force field to study the electric field-driven enzyme catalytic reaction in the enzyme ketosteroid isomerase (KSI). The calculated electric fields induced by the active site of KSI on the carbonyl probe in 19-NT ligand are -108±4.9 MV/cm with AMOEBA. The authors also showed that simulations without mutual polarisation reduced the electric field to −68.08 ± 3.1 MV/cm. The encouraging agreement with the experimental value (i.e., 120-150 MV/cm) for AMOEBA simulations highlights the need for explicit polarisation to capture the changes of the electric fields at the enzyme active site.

Another area of interest is the $O_2$ binding and diffusion in biomolecular systems. $O_2$ is a neutral but highly polarisable molecule and non-polarisable force fields represent its interactions with the environment with van der Waals interactions only.[67] Torabifard and Cisneros compared $O_2$ diffusion in AlkB with the AMBER and AMOEBA force fields.[68] The PMF based on both force fields consistently showed a passive transport of $O_2$ from the surface of the protein to the active site. However, the PMF by AMOEBA shows a larger barrier for diffusion of the co-substrate out of the active site than the non-polarisable force field. It has been suggested that explicit polarisation is crucial to adequately describe the interactions between $O_2$ (neutral albeit highly polarisable) and its environment.

# Ion channels

Electrostatics and polarisation also play an important role in the mechanisms of ion channel gating and conduction.[69] Peng et al. showed that they were able to reproduce the experimental conductance in Gramicidin A with the AMOEBA force field.[70] Sun and Gong[71] modelled the transition in the voltage-gated sodium channel ($Na_V$) from its resting state to the pre-active state using the CHARMM-Drude force field. They were able to show the conformational changes of $Na_V$ from the resting state to the pre-active state. The polarisation of the $\pi$-electrons in Phe56 by the positively charged Arg3 in $Na_V$ was found to stabilise the protein structure when the charged gating residues pass the hydrophobic constriction site during activation. Polarisable force fields have been used to study other ion channels as well.[72,73]

# Membrane permeation

Biological membranes are composed of a bilayer of mixed lipid components with membrane proteins embedded in them. Many cellular signalling and metabolic processes require selective passage of ions or small molecules across the membrane either through non-facilitated permeation through the lipid bilayer or by facilitation by membrane-spanning proteins. These structures inherently possess various electrostatic environments, as ionic or polar headgroups face the interior and exterior solutions to form a water–membrane interface while the interior of the membrane is composed of non-polar saturated and unsaturated lipid tails. As a consequence, molecules permeating through the membrane experience different degrees of polarisation depending on their positions in the membrane.

Induced polarisation can play a significant role in non-facilitated membrane permeation. Small molecules permeating a lipid bilayer cross between the polar aqueous solution, through the ionic water–bilayer interface, and through the non-polar lipid tails in the interior of the bilayer. This range of electrostatic environments results in large shifts in the induced polarisation of permeating solutes. Riahi and Rowley explored these effects in simulations of the permeation of water and hydrogen sulphide through a DPPC lipid bilayer using the CHARMM-Drude polarisable force field.[74] The dipole moment of the permeating water molecule was largest ($<\mu>$=2.5 D) in the aqueous phase where there are strongly-polarising hydrogen bonds with other water molecules. This polarisation decreases as the water molecules enter the bilayer, reaching a minimum at the centre of the membrane, where the dipole moment is ~1.9 D. Hydrogen sulphide shows a similar but less pronounced trend, where the average solute dipole decreases from 1.2 D to 1.0 D (**Figure 3**). This highlights an apparent paradox in the induced polarisation of solutes in condensed phases; highly polarisable molecules such as hydrogen sulphide experience a smaller degree of induced polarisation than the less polarisable water molecules. This reflects that the atomic radii of atoms also increase with their polarisability, so highly polarisable atoms, like S and C, may well be too large to participate in strong, short-range electrostatic interactions that result in a strong induced polarisation effect.

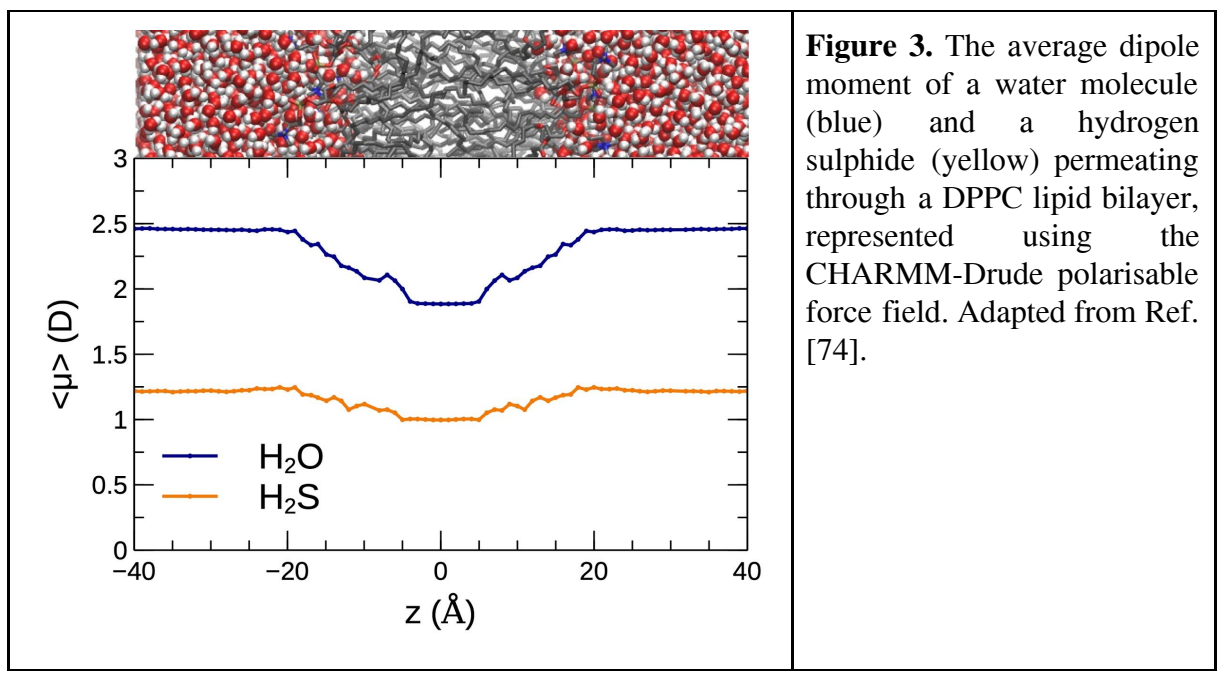

**Figure 3.** The average dipole moment of a water molecule (blue) and a hydrogen sulphide (yellow) permeating through a DPPC lipid bilayer, represented using the CHARMM-Drude polarisable force field. Adapted from Ref. [74].

# QM/MM simulations and computational vibrational spectroscopy

QM/MM MD simulations are powerful methods to study how the environment affects the reactivity or spectroscopic properties of a critical component. An immediate concern is that the enhanced partial charges in additive force fields will create an inconsistent and unbalanced description of the interactions between the QM part and the MM part in combined QM/MM simulations. Polarisable force fields may offer a solution to this issue, and there have been many reports where a QM/MM model was constructed using a polarisable MM model.[75,76] The accuracies of these simulations depend on the QM model, the MM model, and the interactions between QM and MM. König et al. systematically studied the hydration free energies of 12 small molecules with QM/MM simulations with the CHARMM force field and the CHARMM-Drude polarisable force field.[77] Despite the potential for the polarisable model to provide more accurate results, the resulting QM/MM hydration free energies were inferior to purely classical results, with the QM/MM(Drude) predictions being only marginally better than the QM/MM(non-polarisable) results. Ganguly et al.[78] reported the first systematic assessment of a polarisable force field in QM/MM studies of enzymatic reactions. In the cases of the Claisen rearrangement in chorismate mutase and the hydroxylation reaction in p-hydroxybenzoate hydrolase, the authors observed that explicit MM polarisation has moderate effects on activation and reaction (free) energies. They concluded that further validation work is required to establish the best QM/MM-based procedure for handling polarisation effects in enzymatic reactions.

Polarisable force fields have also been applied to the prediction of vibrational spectra, especially where the vibrational models are highly anharmonic in nature or are sensitive to the surrounding electrostatic environment[79]. Semrouni et al.[80] and Thaunay et al.[81,82] applied the AMOEBA force field to calculate vibrational spectra and their temperature-dependence using the Fourier transform of the dipole autocorrelation function. Explicit polarisation could provide improved sensitivity of the spectra to the environment by rigorously including solvent–solute interactions like hydrogen bonds. Furthermore, combined QM and polarisable force field simulations are an attractive method to predict and understand the infrared spectra of molecules in solution and a biomolecular system. [83]

## Conclusions and Outlook

In the past decades, we have witnessed impressive progress in the development of polarisable force fields and their application in biomolecular simulations. This has been enabled by efficient software development and continuous refinement of force field parameters. The applications have provided many new insights into biological processes, where explicit polarisation is crucial. At the same time, more systematic validation is needed to understand and improve some of the limitations in the current models, including both the underlying physical models and their parameterisation. The development of automated and systematic parameterisation techniques is particularly promising.

## Acknowledgements

The authors thank Dr Jing Huang (Westlake University, China), Dr Qiantao Wang (Sichuan University, China) and Mr Koen Visscher (Vrije Universiteit Amsterdam, the Netherlands) for comments on this article. We wish to acknowledge the Australian Government for an Australian International Postgraduate Award scholarship for V.I. DPG acknowledges financial support from the Netherlands Organization for Scientific Research (NWO, VIDI grant 723.012.105). CNR thanks NSERC of Canada for funding through the Discovery Grants program (RGPIN-05795-2016). This research was in part supported under the Australian Research Council's Discovery Project (DP170101773, H.Y.) and National Health and Medical Research Council Project Grant (APP1145760, H.Y.). This article is dedicated to the memory of Prof Herman J.C. Berendsen, one of the founding fathers of biomolecular simulations.

## References and recommended readings

Papers of particular interest, published within the period of review, have been highlighted as:

· **of special interest**

[9] Tinker-HP, a massively parallel molecular dynamics package, has been developed and it enables efficient molecular dynamics simulations with the AMOEBA force field.

[29] The cation-$\pi$ and anion-ring interactions in the CHARMM Drude-2013 force field have been refined. The refined force field leads to a significant improvement in reproducing the geometric properties of the ion-$\pi$ interactions in experimental protein structures.

[53] A protocol was proposed to parameterise a polarisable force field based on quantum mechanical calculations. It showed that combining $C_6$ and $C_8$ attractive terms together with a $C_{11}$ repulsive potential yields satisfying models and this demonstrated that explicit inclusion of higher-order dispersion terms could be viable in polarisable force field development.

[78] The effects of explicit MM polarisation in QM/MM simulations of enzymatic reactions have been systematically studied in two systems. The authors concluded that further validation is required to establish the best QM/MM-based procedure for dealing with the effects of polarisation in enzymatic reactions.

**·· of outstanding interest**

[61] Long-time-scale MD simulations were carried out to evaluate the effect of explicit polarisation in protein simulations. The authors showed that polarisable force fields have better performance in protein structure refinement and sample conformational ensembles of IDPs. They also noted the limitation of the current polarisable force fields in predicting the protein native state.

[63] This study compared the folding free energy landscape of GB1 with non-polarisable and polarisable CHARMM force fields. The additive CHARMM36 force field was found to overstablize the folded hairpin structure of GB1. The CHARMM

Drude-2013 force field predicted the relative stability more accurately, but several areas where this force field could be improved further were also identified. Additionally, they showed that tryptophan fluorescence is insufficient for capturing the full $\beta$-hairpin folding pathway.

[66] The authors reported the effects of conformational dynamics on the fluctuations of electric fields on the carbonyl probe in ketosteroid isomerase (KSI) with the AMOEBA force field. It was found that Asp103 promotes large fluctuations in the electric field along the functional cycle of KSI.